# Data Dissemination Strategies for Emerging Wireless Body-to-Body Networks based Internet of Humans

Dhafer Ben Arbia[12], Muhammad Mahtab Alam[1], Rabah Attia[2], Elyes Ben Hamida[1]

[1] Qatar Mobility Innovations Center (QMIC), Qatar Science and Technology Park (QSTP), PO Box 210531, Doha, Qatar.
[2] SERCOM Lab, Polytechnic School of Tunisia, University of Carthage B.P. 743. 2078 La Marsa. Tunisia.
Email : dhafera@qmic.com, mahtaba@qmic.com, rabah.attia@enit.rnu.tn, elyesb@qmic.com

*Abstract*— With the recent advent of Internet of Humans (IoH), wireless body-to-body networks (WBBNs) are emerging as the fundamental part of this new paradigm. In particular with reference to newly emerging applications, the research trends on data routing and dissemination strategies have gained a great interest in WBBN. In this paper, we present the performance evaluation of the clustered and distributed data dissemination approaches in tactical WBBN. We used a realistic radio-link and biomechanical mobility model for on-body motions, and group mobility model for WBBN to effectively realize rescue and emergency management application scenario. In this regard, we are using the newly proposed IEEE 802.15.6 standard targeted for body area networks. Extensive (IEEE 802.15.6 standard compliance) network level, packet oriented simulations are conducted in WSNet simulator. During the simulations, various payloads, frequencies (narrow-band) and modulation techniques are exploited. We based our performance evaluation on relevant metrics according to the operational requirements for tactical networks such as packet reception ratio, latency, energy consumption and hop count. The results showed a trade-offs between clustered-based and distributed-based dissemination approaches. With regards to packet delay, distributed approach provided the best performance. However, in terms of average packet reception ratio (PRR), clustered-based approach achieves up to 97% reception and remained the best strategy. Whereas, the results of the hop count and energy consumption are almost comparable in both schemes.

I. INTRODUCTION

The internet of things (IoT) is emerging as a key enabling technology for the next generation of inter-connected world. It is based on the concept of unique addressable objects, which can virtually and seamlessly connect to each other any time and everywhere [1]. These objects are equipped with various types of sensors to track useful information about these objects. In addition, these objects can sense the environment and communicate autonomously without human interventions [2]. Further, the data generated by these objects can be very useful to provide innovative solutions for telecommunication, health-care, energy and many other sectors.

The concepts of pervasive and ubiquitous computing are related to the IoT, in the sense that all of these paradigms are enabled by large-scale embedded sensor devices [2]. Recently, this concept is leading towards connected human beings also known as Internet of Humans (IoH) [3], where the personal and wearable devices are always connected enabling seamlessly humans connectivity.

We are interested to exploit the concept of IoH in the context of emergency management especially rescue and critical operations where multiple rescue members are inter-connected to realize effective communication and coordination for timely evacuation. In such scenarios, often the existing infrastructure is either itself damaged or oversaturated. It is envisioned that, wearable wireless sensor networks can aid for not only proper evacuation but also monitor the physiological signals of the team members [4]. One of the fundamental questions for effectively realizing not only on-body but also body-to-body communications is that, which architecture (from the point-of-view of data dissemination) could be most suited? Often the operating conditions and environment are very harsh, dynamic mobility, time varying radio channel; wireless fading and body shadowing adds further hurdles for reliable wireless communication.

To address above mentioned problem, in this paper we present a reliability-aware comparative evaluation of the data disseminated strategies using wearable wireless on-body (WBAN) and body-to-body networks (WBBN). In particular, fully distributed (*i.e.*, any sensor on one body can communicate to any sensor on an other body) and clustered-based dissemination approach (*i.e.*, on-body sensor can only communicate with their own on body coordinator, which effectively being resource rich device enables body-to-body communication) are evaluated. A packet-oriented network simulator is developed to analyze the varying behavior of the two strategies under IEEE 802.15.6 standard compliance. The simulation platform includes enhanced IEEE 802.15.6 proposed channel models by adding space and time variations to have accurate pathloss model. To mimic the application scenario, dynamic mobility patterns are generated based on walking, running, siting and standing movements. Biomechanical mobility model provides deterministic channel variations for WBAN, whereas group mobility is used for WBBN channel. In addition, *carrier sense multiple access collision avoidance* (CSMA/CA) medium access control protocol and *dynamic MANET on demand* (DYMO) routing protocol completes the WBAN and WBBN system. Finally, several performance metrics are considered for comparing the two data disseminating approaches. It includes packet reception ratio (PRR), packet delay, energy consumption and hop counts results.

Rest of the paper is organized as follows; first, a brief application scenario context and motivations are presented in Sec. II, followed by related works in Sec. III. Data dissemination strategies are explained in Sec. IV, and finally simulation results are presented in Sec. V, followed by a conclusion.

## II. CONTEXT AND MOTIVATIONS

This paper presents an ongoing research of the project Critical and Rescue Operations using Wearable Wireless sensor networks (CROW$^2$) [4]. The main objective of the project is to provide ubiquitous wireless communication and monitoring systems for emergency networks in disaster relief. It is anticipated that in emergency, the existing infrastructure network may be damaged (due to disaster itself), out-of-range or over saturated. In this realm, to extend the end-to-end network connectivity, WBANs coordinators can wirelessly interconnect the on-body sensors to external network infrastructures, by exploiting cooperative and multi-hop body-to-body or beyond-WBAN communications. Consequently, the tactical teams led by the commander, should forward their information among each other to reach the team leader, which is connected to a command center or Internet.

Tactical operations, with respect to the different functions, are: military operations, law enforcement, emergency medical and health services, border security, environment protection, fire-fighting, search & rescue and emergency crisis. Other classification according to occurrence, space and time are:

- Routine or day-to-day operations: EMS (Emergency Medical Services: *e.g.* heart attack), fire, law enforcement.
- Multi-discipline, multi-jurisdiction: for example, explosion in chemical or nuclear plant.

Planning, triggering and conducting tactical operations depends on varied circumstances such as type of threats (*e.g.* natural disaster, war, *etc.*), location (*e.g.* land: rural, urban, mountain; sea: ocean, coastline; air; underground), weather conditions, *etc.* These circumstances are closely related to the mission requirements and the effective communication is one of the relevant challenges in tactical operations. Real-time information (*i.e.* text, images, videos) transfer is needed before, during and even after the operation. This information is required in different tiers. For example, involved personal are concerned at the first level, then, the on-the-field teams' leaders and finally the command center. Hence, communication network must cover all these cited commanding and executing levels. However, it is anticipated that, during and after a crisis or a disaster, existing networking infrastructures can be either completely damaged or oversaturated. Therefore, a tactical network should be instantly deployable to connect the crisis area (*i.e.* engaged staff on the field) to a distant command center and internet. In this framework, we identify two types of communication regarding the engaged personals: *(i)* On-Body communications: data flow coming from on-body nodes (*i.e.* sensors, GPS, *etc.*) going to the coordinator (*i.e.* sink node) and vice-versa. *(ii)* Body-to-Body communications: data flow going from Body to another Body. Our interest in this work is to consider two dissimilar data dissemination strategies while routing data from simple On-Body nodes to the team leader's coordinator which is connected to the large network and eventually Internet. We consider cluster-based and distributed dissemination strategies which are detailed in Section IV.

## III. RELATED WORKS

In the literature, diverse data dissemination protocols have been proposed for Wireless Sensor Networks (WSNs) [5]. WSN is composed by a certain number of sensor devices distributed on an area of interest. Sensor devices are severely constrained in terms of memory, computation capabilities, wireless range and battery power. Sensors (*i.e.*, source-node) sense the environment physical measurements and send them towards a sink (*i.e.*, destination-node). The sensing process could be either triggered by the source-node (*i.e.*, through *periodic sensing*), or depending on the events (*i.e.*, *Event driven*) or requested by the sink (*i.e.*, *Query Based*). Data dissemination strategies for WSN are adopted recently for the Wireless Body-to-Body Networks (WBBNs) with major restrictions [6] [7].

Based on different strategies, sensed data is disseminated towards the sink node. These strategies are classified with respect to: *(i)* type of disseminated data, *(ii)* depending on the destination(s), where both uses the concept of virtual infrastructure [8]. Further for *(i)*, there are three categories: *data dissemination* (where the sensed measurements are disseminated), *meta-data dissemination* (where the sensed measurements are stored locally and a meta-data is disseminated), and *Sink location dissemination* (where the locations are stored into nodes information, and then data is disseminated depending on events). For *(ii)*, dissemination strategies are categorized as: *single node* (disseminated information is stored in one node), *out-of-group nodes* (the information is disseminated out of a defined group of nodes), *a set of nodes* (information is depicted into a set of nodes). Most known data dissemination protocols in WSNs are Directed Diffusion (DD), Geographic Hash Table (GHT), Two-Tier Data Dissemination (TTDD), Railroad, Locators, *etc.*

It is important to note that, in Emergency and Critical networks, WSN could be a part of a WBAN. A WBAN is a set of miniaturized devices (*i.e.* sensors, GPS, RFid tags/readers) wirelessly interconnected and attached (or implanted) into body (human, animal, *etc.*). All these devices are connected with a sink node (*i.e.* coordinator). Despite the fact that some of the above WSNs protocols were evaluated in a single WBAN context, WBAN still have considerable particularities against WSN [9]. First, mobility in WBAN is more important than WSN (*i.e.* WSN are considered stationary) therefore, link failure consideration among devices is relevant. Second, in critical operations, devices battery lifetime used in WBAN, is not a crucial requirement (during operations batteries could be replaced or recharged) instead of scattered sensors (in case of WSN) where battery must operate for long time (few years). These particularities impact requirements of data dissemination protocols. Classic data dissemination strategies within single WBAN were based on links lifetime. However, recent dissemination mechanisms tend to be more opportunistic and posture-aware due to the high WBAN dynamic variations especially in tactical operations. Opportunistic dissemination techniques prove energy preservation and network lifetime increase [10].

Moreover, probabilistic routing protocols use the historical link quality estimation and the inertial sensor data to make the best relaying decision.

Further researches consist on evaluating Ad hoc routing protocols in a scale of single WBAN. Asogwa *et al.* in [11] evaluated Ad hoc On-demand Distance Vector (*i.e.* AODV), Dynamic Source Routing (*i.e.* DSR) and Destination-Sequenced Distance-Vector (*i.e.* DSDV) routing techniques. The obtained results showed that AODV and DSR have good reliability and performed much better in terms of energy efficiency. Likewise, according to Murthy *et al.* [12], AODV is the most efficient routing protocol for intra-BAN in terms of energy efficiency and QoS.

At an extended level, Ad hoc routing techniques were also used to cover Body-to-Body communications [13]. Even more recent, an interesting layer-2 (*i.e.*, MAC Layer) data forwarding strategy proposed by Kolios *et al.* with reference to a specific Emergency Ad hoc Network (*i.e.*, ERN) [14]. Explore and Exploit (*i.e.*, EnE) data dissemination strategy is based on new topology-related metric, Local Centrality (*i.e.*, LC). LC computes a node importance rank that classifies the nodes based on their topological properties. Alert Messages (*i.e.*, AM) will be disseminated through the nodes with highest LC. Indeed, no routing calculations, building and maintenance is needed, thus, no network protocol is implemented. LC information is stored into the layer-2 headers. According to the authors, EnE requires trivial communication overhead and includes smart forwarders selection.

To conclude, existing data dissemination approaches in WBBN are primarily based on the operational context (*i.e.*, use cases: critical, emergency, delay-tolerant, *etc.*), next, on the type of the data to disseminate (*i.e.*, location, data, meta-data). MANET are evidently evaluated in WBBN, however, data dissemination strategies depending on operational requirements are not yet investigated. An important operational requirement in tactical operations consist on; the team leader has to be able to receive, follow and feedback the operation commanding center of all the information provided by his team. For this, clustered or distributed approaches are investigated in the following section.

## IV. DATA DISSEMINATION STRATEGIES

During the last decade, most of the studies were focused on to the feasibility of the MANETs in tactical networks. This tendency is justified by the fact that the tactical operations happens in rural and populated areas where networking infrastructures are either absent or shattered, which comply with the specifications of the tactical operations. Furthermore, due to its flexibility to topology changes and its multi-hop routing aspect, Mobile Ad hoc networks are an interesting candidate to be investigated in the tactical WBANs. In this regard, one of the experimental works evaluated MANETs in rescue and critical operations [15]. However, we introduce the following network architecture which is based on the principle that each team member has to send all the information as *One-Way-Converge-Cast* traffic towards the unique team leader.

*1) Network Architecture*

As mentioned in Section III, proposed approaches for data dissemination were either classified by type of the disseminated information or based on nodes status (energy, connectivity, *etc.*). The objective of this paper is to evaluate the performance of two data dissemination strategies (clustered and distributed) with specific simulation setup detailed in the next section. The disseminated information towards the team leader could reach it in two different ways:

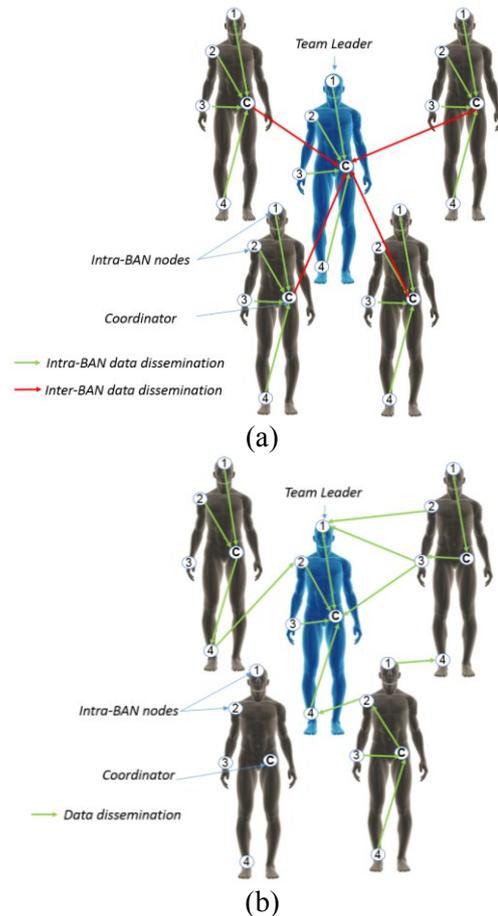

Figure 1 (a): Clustered and (b) Distributed data dissemination strategy

- *Clustered Data Dissemination (CDD)*: From one-WBAN nodes to their coordinator (*i.e.*, embedded on the same WBAN), and then from that coordinator to the adjacent coordinators until reaching the team leader's coordinator (*i.e.*, a coordinator is a sink node responsible for gathering data from the other On-Body nodes, it is characterized by powerful capabilities comparing to the other nodes). Thus, each WBAN could operate in a single frequency. Fig. 1a shows in the green color, a communication link between the sensor nodes and a coordinating node during the data dissemination process, whereas, in red color, the *Coordinator-to-Coordinator* data disseminations are highlighted.

- *Distributed Data Dissemination (DDD)*: From one-BAN nodes to the any reachable adjacent nodes (simple node or coordinator), until reaching the team leader's coordinator. Consequently, all nodes need to share same frequency which could raise an interference issue. Fig. 1b depicts the distributed data dissemination, any node could send its data to any node, however, the final destination is always the coordinator of the Team Leader.

*2) Routing Protocol*

Without loss of generality, and as discussed in Section III, AODV was found as the most efficient routing protocol in intra-BAN communications. However, despite the latest trends to evaluate Ad hoc routing techniques in critical and emergency environments, few studies have been conducted to evaluate these strategies with realistic mobility model. In one of our recent work, we evaluated four routing protocols (*i.e.*, proactive, reactive, gradient-based and geographical-based) in emergency and critical context. So, we found that AODV version 2 showed a considerable behavior in this context. In contrast, this work does not assess routing protocols but the way data is disseminated in intra and inter-WBANs. We consider AODV version 2 as the routing protocol in our following investigations.

AODV version 2 (*i.e.*, DYMO) [16] routing protocol is considered as an enhancement of AODV, with recourse to some of the features of DSR. Indeed, DYMO uses 'path accumulation' from DSR and removes unnecessary Route Reply (RREP), precursor lists and Hello messages (*aka* Route exploration messages). From AODV, DYMO keeps sequence number, hop count and RERR. DYMO has two main operations: route discovery and route management. In DYMO routing protocol, the route discovery process starts with the RREQ (if no route to destination exists in the source routing table), then, each time the RREQ is forwarded throughout the network, forwarding node will attach its address to the RREQ message. Once the destination reached, The RREP will be sent in unicast to the source node following the accumulated path. DYMO is an energy efficient protocol, then if one node has low energy it does not participate in the route discovery process since it may be disconnected until the RREP is sent back. Obviously, while sending data, if the link is break or the destination node is no more available, the intermediate node multicasts a RERR to only concerned nodes by the failed links. Upon the reception of the RERR, the routing table entry containing the unavailable node will be deleted. Hereafter, a new route discovery will be initiated when a destination is needed.

V. SIMULATION RESULTS

In this section, we investigate the performance behavior of the previously discussed data dissemination strategies and routing protocol under realistic WBBN channel and mobility modeling. The detailed simulation parameters and models are firstly described, followed by the obtained simulation results in terms of packet reception ratio, energy consumption, communication delays and routes hop count.

*1) Radio Link and Mobility Modeling*

Accurate radio link and mobility modeling is key to the system performance of wearable tactical network introduced in previous section. Our radio-link modeling metric is based on SINR (signal-to-interference-noise-ratio), which takes into account the mutual interference from multiple WBANs [6]. This metric rely on accurate path loss calculations using enhanced IEEE 802.15.6 channels models [17]. Then, *bit error rate* is calculated based on the specific modulation schemes (*i.e.*, DQPSK and DBPSK) proposed in the IEEE 802.15.6 standard, followed by the evaluation of *packet error rate* (PER) [6].

In WBANs there are different mobility patterns depending upon the posture positions during sitting, standing, walking, running swimming *etc.*, scenarios. In addition, body shadowing, orientation and rotations make the radio-link consistently time-varying. Our modeling methodology is based on real-time mobility traces from the motion captured system which provides diverse mobility patterns such as walking, sitting, standing and running. These mobility patterns coincide with our application scenario as explained in Sec. II, which are imported in a packet-oriented network simulator (called WSNet) for complete system design and performance analysis. Further, we have developed bio-mechanical models (for on-body) communication and group mobility model (for body-to-body) communications which reflect and satisfy our application context. The detailed steps of the biomechanical modeling and transformation are explained in [17]. In this paper we have considered three levels of hierarchy as shown in Fig. 2. At the top level, there are 12 bodies (WBANs), then, 3 WBANs form a small group (for inter-WBAN mobility) and finally each body consists of five on-body nodes. Concerning the separation distance between these WBANs, the WBANs inside a small WBANs group are separated by 8 meters, whereas, 20 meters separation is considered between the groups. Five on-body nodes are placed as; head (node 1), right shoulder (node 2), right wrist (node 3), stomach (node 0), and right ankle (node 4).

*B. Simulations Setup*

We consider the reference scenario of Section IV, in which twelve wearable body area networks (WBANs) are evolving in a tactical group formation. The bodies' mobility patterns include sitting, standing, walking (*i.e.*, 0.5m/s) and running (*i.e.*, 3m/s). We consider two different nodes architectures based on the aforementioned data dissemination strategies. In the distributed data dissemination strategy, all on-body sensors (including WBANs coordinators) are running on top of an

IEEE 802.15.6 compliant MAC and PHY layers, with the power consumption characteristics of the CC2420 RF transceiver [18].

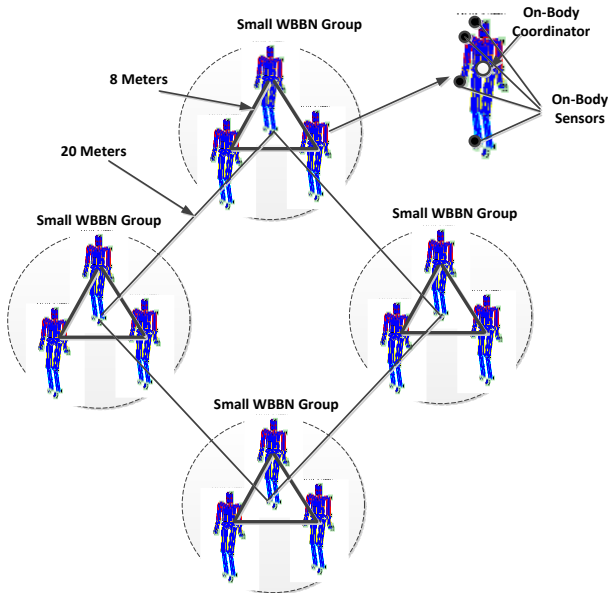

Figure 2. Tactical Wireless Body-to-Body Network Scenario.

Concerning the PHY layer parameters, the transmission power was set to 0dBm, two frequencies were evaluated (*i.e.*, 2450MHz and 900MHz), and for each frequency two different data rates are considered, *i.e.* 101.2Kbps (DBPSK) and 404.8Kbps (DQPSK) for 900MHz, and 121.4Kbps (DBPSK) and 971.4Kbps (DQPSK) for 2450 MHz. The MAC layer is based on the CSMA/CA protocol with immediate acknowledgement, where all WBANs nodes are operating under the same channel frequency. On top of the MAC layer, the AODV version 2 (DYMO) was implemented with a neighbor discovery frequency of 3s and a timeout of 9s. Finally, a Constant Bit Rate (CBR) application is generating data traffic on all WBANs nodes using different data payloads (*i.e.* from 16Bytes to 256 Bytes) and frequencies (*i.e.* 250ms, 500ms and 1s).

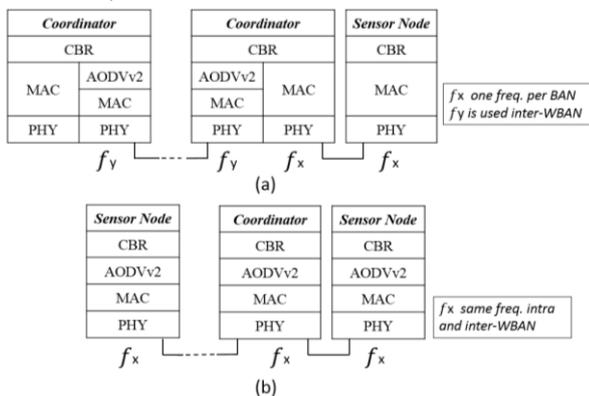

Figure 3 (a) Clustered approach where one frequency is used per BAN and a different frequency is used for inter-WBAN. (b) Distributed approach where same frequency is used from any node to any node (even coordinator)

Fig. 3 shows the node architectures (for both sensors and coordinator) under distributed and clustered approaches. In the clustered data dissemination strategy, each WBAN coordinator device is based on a multi-standard communication stack, where one MAC/PHY interface is used to communicate with the on-body sensors through a dedicated channel frequency (each WBAN uses a different channel frequency to avoid interferences with other WBANs), whereas the second MAC/PHY interface is used to communicate with the surrounding WBANs coordinator using a same channel frequency. In this case, the communication between the on-body sensors and their coordinator is performed using CSMA/CA, whereas AODV version 2 is only implemented at the coordinator node to discover the surrounding coordinator devices from the other WBANs, and to route the collected data to the WBANs group leader. We considered 10 iterations for each simulation scenario, and the 95% confidence intervals were computed and reported in the below simulation results.

*C. Results*

*1) Packet Reception Ratio (PRR)*

Fig. 4, shows the results of average PRR against varying payload (*i.e.*, from 16-to-1024 bytes) transmitted per second for the application layer by each of the four sensors and coordinator connected on the body. In addition, 900 MHz and 2450 MHz narrow-band frequencies are utilized with lowest and highest data rates as specified earlier. In general it can be seen that clustered-based approach achieves much better PRR under both frequencies with DQPSK (*i.e.*, highest rate). Whereas, DBPSK (*i.e.*, lowest data rate), in distributed approach achieves the lowest performance in both frequency under all payloads variations. Further, it can be seen in both Fig. 4a and Fig. 4b, that there is a gradual decrease in PRR performance with an increase in the payloads.

In specific, with low payload, clustered approach achieves almost 97% PRR; however, the performance degrades relatively more with the higher payloads especially when operating at 900 MHz frequency. The best performance of the clustered-based strategy at the maximum payload (*i.e.*, 256 bytes) is with DQPSK at 2450 MHz, where the PRR drops up to 75%. On the other hand, distributed approach with the highest rate is comparable with clustered approach (lowest rate) at 900 MHz, though it performs slightly better in 2450 MHz frequency. However, the results are always below 80% PRR even at 2 bytes of payload.

*2) Latency*

As per the packet delay performance, Fig. 5 shows the average of packet transmission delay. Payload is varied as from 16-to-1024 bytes are transmitted per second. As well, 900 and 2450 MHz are the utilized frequencies. Generally, the results of the delay are inter-related with PRR, if PRR is higher then, delay will be lower due to higher successful transmissions and lower retransmissions. It is clear that both clustered and distributed-based approaches have similar behavior with DQPSK with different variation of the payload and frequencies. Best average delay is given by the distributed-based approach with 64 bytes payload at both utilized frequencies. Accordingly to the PRR, worst performance is noticed for distributed-based dissemination strategy for all payload values and frequencies.

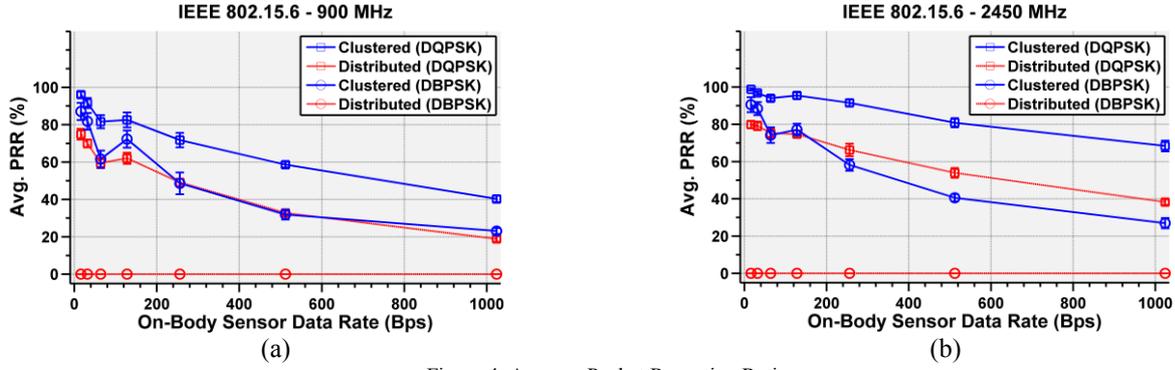

(a)                                (b)

Figure 4. Average Packet Reception Ratio.

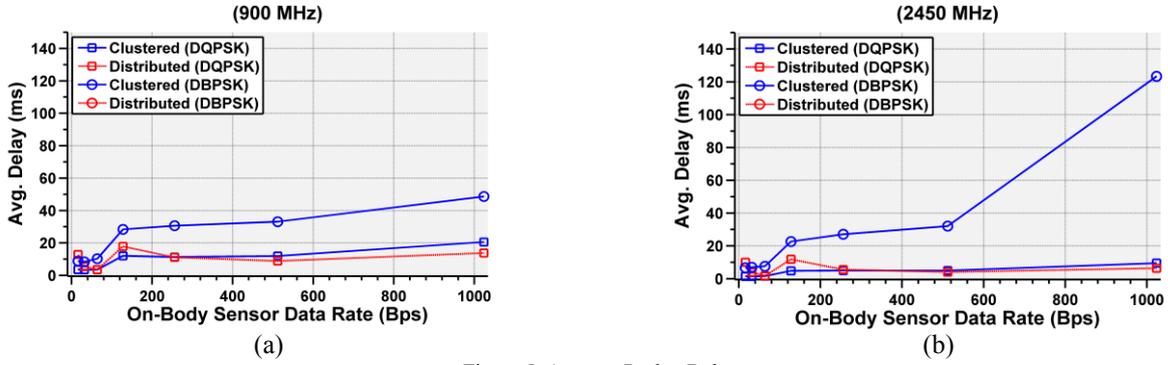

(a)                                (b)

Figure 5. Average Packet Delay.

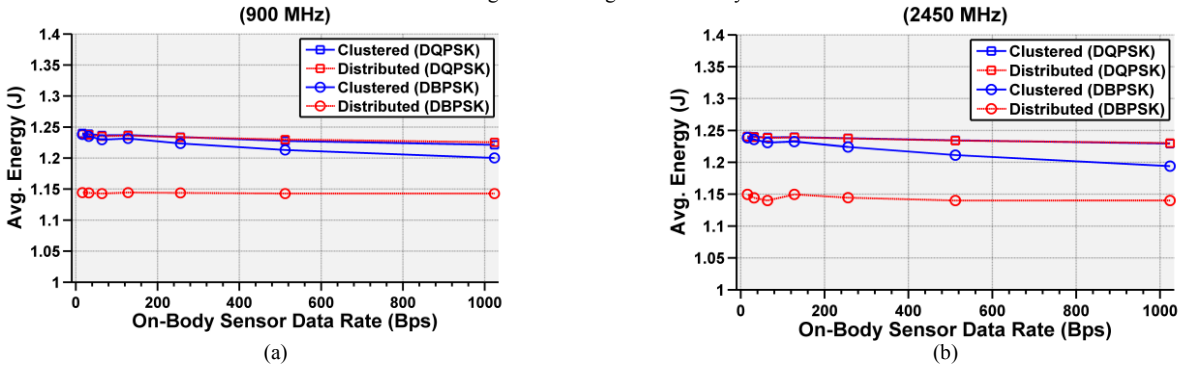

(a)                                (b)

Figure 6. Average Energy Consumption.

Specifically, with low payload and high rate (*i.e.* DQPSK), distributed and clustered-based approaches latency is interesting with a delay under 10ms. In contrast, for distributed dissemination strategy, delay is infinite with DPSK, which is relatively expected from the PRR average (≈ 0%). DBPSK in clustered-based approach, has a linear increase to reach 50ms with highest payload. (*i.e.* 1024 bytes).

*3) Energy Consumption*

Concerning the energy consumption, Fig. 6 shows the energy consumption for clustered and distributed data dissemination approaches with low and high rate (*i.e.* DBPSK and DQPSK). The energy consumption is shown with two graphs respectively for 900 and 2450 MHz as utilized frequencies. The energy consumption for each transmitted packet is calculated as follows,

$$E_{packet} = T_{packet} \times 3_{Volts} \times I_{mA}$$

where, $T_{packet}$ is the duration in ms which is based on the effective packet length (including all the *PHY and MAC headers* [19]).

It can be seen that in general similarly for both utilized frequencies, with DQPSK (*i.e.* highest rate) energy consumption follows the same curve for the two investigated data dissemination strategies. Distributed approach with DBPSK for both frequencies (900 and 2450 MHz) shows the lowest values for energy consumption, this is explained by the null PRR average depicted in Fig. 4. Actually, there is no packets sent in this case (Distributed with DBPSK), so the energy consumption will be consequently the lowest. Clearly, clustered approach with DBPSK consumes slightly low energy compared to DQPSK for both dissemination approaches. However, even though clustered approach with DBPSK is performing with lowest energy consumption, according to the delay discussed based on Fig. 5, is not the most performant approach. Finally, DQPSK digital modulation has the same energy consumption behavior for both dissemination strategies.

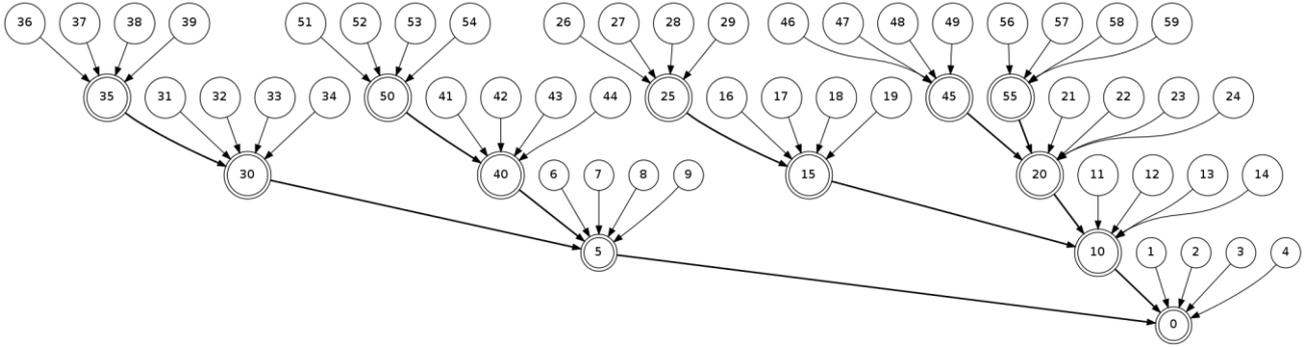

Figure 7. Network topology obtained with the clustered routing approach (2450Mhz, DQPSK, and Payload of 16 bytes).

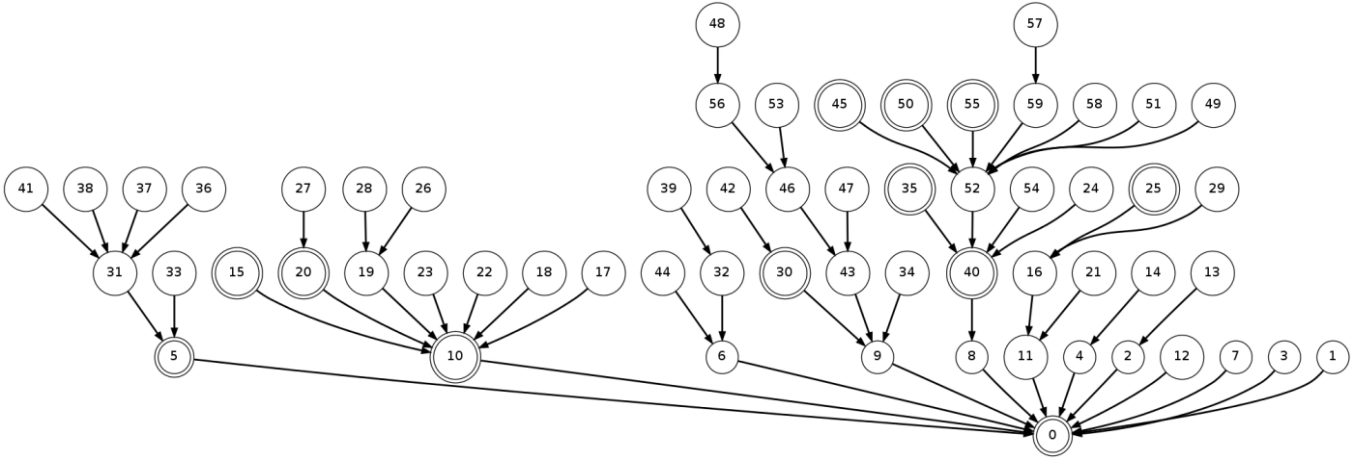

Figure 8. Network topology obtained with the distributed routing approach (2450Mhz, DQPSK, and Payload of 16 bytes).

*4) Hop Count*

TABLE I. shows the hop count for different data dissemination approaches with high and low rate and the utilized frequencies detailed above. Hop count is an important metric in tactical networks. Hence, it is considered as the relevant routing decisive parameter. In general, with the digital modulation DQPSK and both utilized frequencies, clustered and distributed dissemination approaches have almost the same hop count average (*i.e.*, from 2,24 to 2,48). With DBPSK, clustered dissemination approach has the same hop count average with both frequencies. Distributed approach with DBPSK with both utilized frequencies is not considered based on the PRR average. Specifically, digital modulation DQPSK is most appropriate for clustered and distributed dissemination approaches in terms of hop count.

TABLE I. HOPS COUNT STATISTICS (COMPUTED ACROSS ALL DATA PAYLOADS AND ITERATIONS)

| PHY Layer | Routing Layer | Hops Count | | |
|---|---|---|---|---|
| | | Min | Average | Max |
| 2450 Mhz + DQPSK | Distributed | 1 | 2.44 | 7 |
| | Clustered | 1 | 2.24 | 6 |
| 2450 Mhz + DBPSK | Distributed | N/A | N/A | N/A |
| | Clustered | 1 | 1.21 | 3 |
| 900 Mhz + DQPSK | Distributed | 1 | 2.48 | 7 |
| | Clustered | 1 | 2.25 | 5 |
| 900 Mhz + DBPSK | Distributed | N/A | N/A | N/A |
| | Clustered | 1 | 1.26 | 4 |

To conclude, Fig. 7 and Fig. 8 depict the network topology obtained with the clustered and distributed dissemination approaches (2450 MHz, DQPSK, and Payload of 16 bytes). Figures show clearly that number of hops for most of the nodes is much better with the distributed approach (Fig. 7). However, the PRR average (for 2450 MHz, DQPSK and Payload of 16 bytes) is more important with clustered approach (*i.e.*, 89%). So, there is a clear trade-off between both data dissemination approaches.

## VI. CONCLUSIONS

This paper provides a comparison between clustered and distributed data dissemination strategies in Wireless Body-to-Body tactical networks. Currently there are very few studies which have evaluated the performance of such emerging networks especially in the context of emergency management and rescue operations. We evaluated the performance of WBBN architecture using realistic radio-link and mobility models, in addition, enhanced IEEE 802.15.6 standard proposed channel models are utilized from our recent ongoing research. Extensive simulations are conducted to evaluate the performance of data dissemination strategies in WBBN using two modulation techniques (DQPSK and DBPSK) and with two different frequencies (900 and 2450 MHz). Our results showed that there is a trade-off between the two approaches. For example, regarding PRR, clustered based data dissemination approach performed much better than the distributed under varying payloads with both 900 MHz and 2450 MHz operating frequencies. In particular, using DQPSK modulation (which provides highest data rates) achieves up to 97% PRR. Whereas, concerning the delay, although clustered-based dissemination strategy using DQPSK provides a considerable options, but distributed approach maintain lowest delay in all configurations and is considered as the best approach. In terms of energy consumption, with DQPSK clustered and distributed-based approaches have similar evolution for different given payloads. Hence, it is concluded that for small tactical teams (*i.e.*, 12 member), with DQPSK modulation technique and for the two utilized frequencies, clustered-based data dissemination approach is appropriate for Body-to-Body communications.

ACKNOWLEDGMENT

This publication was made possible by NPRP grant # [6-1508-2-616] from the Qatar National Research Fund (a member of Qatar Foundation). The statements made herein are solely the responsibility of the authors.

REFERENCES

[1] K. Ashton, "That 'Internet of Things' Thing," *RFID Journals*, 2009.

[2] C., N. Aggarwal, A. Ashish and Sheth, "THE INTERNET OF THINGS: A SURVEY FROM THE DATA-CENTRIC PERSPECTIVE," in *Managing and Mining Sensor Data*, Springer, 2013, pp. 384-428.

[3] A. and Bloom, *Trends: The Internet of Humans (not Things)—Sensors, Health, Fitness, & Healthcare*, Pivotal, 2015.

[4] E. Ben Hamida, M. M. Alam, M. Maman, B. Denis and R. D'Errico, "Wearable Body-to-Body Networks for Critical and Rescue Operations - The CROW² Project," in *IEEE 25th Annual International Symposium on Personal, Indoor, and Mobile Radio Communications (IEEE PIMRC 2014), Workshop on The Convergence of Wireless Technologies for Personalized Healthcare*, Washington DC, 2014.

[5] E. B. Hamida and G. Chelius, "Analytical Evaluation of Virtual Infrastructures for Data Dissemination in Wireless Sensor Networks with Mobile Sink," in *Proceedings of the First ACM Workshop on Sensor and Actor Networks (SANET)*, Montreal, Quebec, Canada, 2007.

[6] M. M. Alam and E. B. Hamida, "Interference Mitigation and Coexistence Strategies in IEEE 802.15.6 based Wearable Body-to-Body Networks," in *10th International Conference on Cognitive Radio Oriented Wireless Networks (CrownCom 2015)*, 2015.

[7] M. M. Alam and E. Ben Hamida, "Surveying Wearable Human Assistive Technology for Life and Safety Critical Applications: Standards, Challenges and Opportunities," *Sensors*, vol. 14, no. 5, pp. 9153 - 9209, 2014.

[8] E. Hamida and G. Chelius, "Strategies for data dissemination to mobile sinks in wireless sensor networks," *IEEE Wireless Communications*, vol. 15, no. 6, pp. 31-37, 2008.

[9] D. He, S. Chan, Y. Zhang and H. Yang, "Lightweight and Confidential Data Discovery and Dissemination for Wireless Body Area Networks," *IEEE Journal of Biomedical and Health Informatics*, vol. 18, no. 2, pp. 440-448, 2014.

[10] E. Ben Hamida, M. Alam, M. Maman and B. Denis, "Short-term link quality estimation for Opportunistic and Mobility Aware Routing in wearable body sensors networks," in *IEEE 10th International Conference on Wireless and Mobile Computing, Networking and Communications (WiMob)*, 2014.

[11] C. Asogwa, X. Zhang, D. Xiao and A. Hamed, "Experimental Analysis of AODV, DSR and DSDV Protocols Based on Wireless Body Area Network," *Internet of Things - Communications in Computer and Information Science*, vol. 312, pp. 183-191, 2012.

[12] J. K. Murthy, P. Thimmappa and V. Sambasiva Rao, "Investigations on the Routing Protocols for Wireless Body Area Networks," in *Proceedings of International Conference on Advances in Computing*, 2012.

[13] C. Bohannan, L. Zhang, J. Tang, R. S. Wolff, S. Wan, N. Gurdasani and D. Galarus, "QoS Enhancement and Performance Evaluation of ad-hoc routing protocols for rural public safety," in *International Conference on Communications*, Dresden, Germany, 2009.

[14] P. Kolios, A. Pitsillides, O. Mokryn and K. Papadaki, "Qualifying explore and exploit for efficient data dissemination in emergency adhoc networks," in *IEEE International Conference on Pervasive Computing and Communications Workshops (PERCOM Workshops)*, 2014.

[15] W. Richard , T. Jian , D. Galarus and L. David , "Ad Hoc Routing for Rural Public Safety," U.S. Department of Homeland Security-Science and Technology Directorate-Office for Interoperability and Compatibility, Homeland, USA, 2008.

[16] C. Perkins, S. Ratliff, J. Dowdell, L. Steenbrink and V. Mercieca, "Dynamic MANET On-demand (AODVv2) Routing - draft-ietf-manet-aodvv2-10," IETF - Mobile Ad hoc Networks Working Group, 2015.

[17] M. M. Alam and E. Ben-Hamida, "Towards Accurate Mobility and Radio Link Modeling for IEEE 802.15.6 Wearable Body Sensor Networks," in *IEEE 10th International Conference on Wireless and Mobile Computing, Networking and Communications (WiMob)*, Larnaca, 2014.

[18] "cc2420 Manual," Chipcon Products from Texas Instruments, 2007. [Online]. Available: http://ti.com/lit/ds/symlink/cc2420.pdf.

[19] M. Alam and E. Ben Hamida, "Performance evaluation of IEEE 802.15.6 MAC for Wearable Body Sensor Networks using a Space-Time dependent radio link model," in *IEEE/ACS 11th International Conference on Computer Systems and Applications (AICCSA)*, 2014.